\documentclass[%
reprint,
amsmath,amssymb,
aps,
showkeys,
]{revtex4-2}

\usepackage{graphicx}% Include figure files
\usepackage{dcolumn}% Align table columns on decimal point
\usepackage{bm}% bold math
\usepackage{subcaption}
\usepackage{ragged2e}
\begin{document}

%\preprint{APS/123-QED}

\title{Revealing more on complex energy landscapes by passing less local information: Cavity approach with trust region in non-convex optimization problems}% Force line breaks with \\

\author{Zhiqi Xu}\author{Chi Ho Yeung}
\email{chyeung@eduhk.hk}
\affiliation{%
Department of Science and Environmental Studies, The Education University of Hong Kong, Tai Po, Hong Kong, China
}%

%\date{\today}

\begin{abstract}
Energy landscapes of non-convex optimization problems are high-dimensional surfaces that are difficult to reveal, visualize, or analyze. Message-passing algorithms, or equivalently cavity approaches in statistical physics, may fail to identify local minima as messages do not converge owing to the ruggedness of the energy landscapes. Here we aim to reveal the characteristics of these complex energy landscapes by limiting the amount of information passed in local messages, improving the convergence of messages for local minima. 
By studying a routing optimization problem with its convexity governed by a single parameter, we introduce a cavity message-passing algorithm with a trust region, and sample ensembles of converged local minima of rugged energy landscapes across many independent realizations of the same problem instances. We observe three regimes with different characteristics of the energy landscapes: (1) a maximally rugged regime with a peak in the diversity of low-energy minima; (2) an intermediate regime with a hierarchical multi-cluster organization of low-energy solutions; and (3) a smooth regime dominated by a single minimum. Additional tests show that the converged energies are largely insensitive to the size of the trust region if the size is small or moderate, demonstrating the robustness of the proposed approach in identifying characteristics of rugged energy landscapes.
\end{abstract}


\maketitle

%\tableofcontents

\section{Introduction}
\label{sec:level1}

The macroscopic properties of physical systems, such as phase transitions and relaxation dynamics, are largely determined by the geometry of their energy landscapes. Specifically, the number of metastable states and the heights of energy barriers are important features of the energy landscape. Spin glasses, which provide representative physical models with a rugged energy landscape, have significantly advanced our understanding of computational complexity in disordered systems and non-convex optimization problems~\cite{mezard1987spin}. In recent years, the physical insights in systems with rugged energy landscapes have been applied in fields such as computer science, biology, and sociology, demonstrating that the concept of energy landscapes can transcend traditional physical systems to offer profound perspectives for many complex network systems~\cite{onuchic1997theory,marvel2009energy,wales2018exploring}.

Although the geometry of an energy landscape is central to understanding phenomena such as metastability, barrier crossing, and dynamical trapping, direct visualization is difficult because the landscape is high dimensional. Existing approaches capture only partial aspects of the underlying structure. A classical approach is the disconnectivity graph, which summarizes how attraction basins split hierarchically as the energy threshold is lowered and provides a compact view of basin connectivity~\cite{becker1997topology,wales2018exploring}. Another common approach is dimensionality reduction, including multidimensional scaling (MDS), which aims to preserve pairwise distances among states in a low-dimensional embedding although such embeddings do not by themselves explicitly encode barrier heights or transition pathways~\cite{mead1992mds}. Another line of work focuses on the organization of low-energy solutions, such as the hierarchical structure of spin-glass ground states revealed by clustering analyses~\cite{hed2001spin}. More recently, complete visualizations of rugged energy landscapes have been proposed for small problems, highlighting both the feasibility and the challenges of revealing full landscape structure in complex systems~\cite{po2024complete}. Taken together, these studies suggest that basin hierarchy, pairwise similarity, and dynamical connectivity are related but not equivalent features, and that no single representation fully captures a high-dimensional rugged landscape.

On the other hand, the cavity approach was originally developed in the study of disordered systems such as spin glasses, and later gave rise to distributed message-passing frameworks that bridge statistical physics, probabilistic inference, and optimization problems~\cite{mezard1987spin,mezard2009information,mezard2016cavity}. In the conventional cavity framework, one considers a node in a graph in which one of its neighboring nodes is removed. The cavity fields (i.e.\ messages) are computed from the cavity fields of the remaining neighboring nodes. For finite discrete variables, cavity messages can often be parametrized by a small number of scalars, whereas for continuous or higher-dimensional variables they are typically functional objects~\cite{mezard2009information,mezard2016cavity}. By passing cavity fields between neighboring nodes, as in other message-passing algorithms, approximately optimal or locally stable solutions can be identified from the converged cavity fields. Cavity-inspired message-passing algorithms have also been adapted to networking and routing problems on sparse graphs~\cite{PhysRevLett.108.208701,yeung2013efficient,yeung2013networking}.

However, cavity fields may not converge, especially when the energy landscapes are characterized by multiple local minima, such as those in non-convex optimization problems. In this study, we introduce a variant of the cavity approach in which only part of the whole functional cavity field is passed between neighbors. In a manner analogous to a trust region, only a local region of messages around a reference point is passed, while the size of the trust region and the reference point are continuously updated until convergence. In our case, trust-region cavity fields are the local expansion of the cavity field around the reference point. We apply this approach to a routing coordination problem in which a single model parameter controls the transition from convex to non-convex optimization landscapes. Remarkably, we can reveal more information on the energy landscapes by applying a smaller trust region in cavity fields, i.e., passing less information. We further conduct a hierarchical clustering analysis of the identified local minima, which reveals different regimes characterized by distinct clustering structures, as well as similarities to the clustering of local minima in spin glasses. In summary, this work introduces a trust-region cavity approach to reveal the energy landscapes of non-convex optimization problems. 

\section{Model}
\label{sec:level2}

Route coordination is an important problem underpinning modern communication and transportation infrastructures. The objective is to coordinate multiple paths over a given network to minimize a global cost function, such as the total delay or congestion~\cite{wang2008overview}. Traditional approaches in computer science often rely on heuristic or online learning algorithms, focusing on the core task of finding optimal or near-optimal path configurations~\cite{medhi2017network}. However, conventional optimization approaches largely overlook the global structure of the solution space. A thorough understanding of the global structure of this energy landscape—including its non-convexity, the potential emergence of a glass-like regime, and its relationship with the network topology—is crucial for assessing algorithmic efficiency, network robustness, and the system's response to external perturbations~\cite{schon2024energy}.

Here, we consider an undirected sparse random graph of $N$ nodes, indexed $i = 1, 2, \dots, N$. Each node is randomly connected to a fixed number $k$ of neighbors, with the set of neighbors of node $i$ denoted by $\mathcal{N}_i$. A fraction $f_p$ ($0 < f_p < 1$) of nodes are designated as \emph{senders}, which serve as the origins of communications directed to a common \emph{receiver} node in the network. We denote by $\sigma^{(s\to r)}_{ij}\in\{0,1\}$ whether the communication sent by sender $s$ to receiver $r$ uses the edge from $i\to j$.

Following Ref.~\cite{PhysRevLett.108.208701}, we work with the net flow
\begin{equation}
I_{ij}\equiv \sum_{s}\big(\sigma^{(s\to r)}_{ij}-\sigma^{(s\to r)}_{ji}\big),\qquad I_{ji}=-I_{ij}.
\end{equation}
To ensure that communications from all senders arrive at the common receiver, flow conservation is imposed on every non-receiver node $i\neq r$ via
\begin{equation}
R_i=\Lambda_i+\sum_{j\in \mathcal{N}_i} I_{ji}=0,\qquad \forall i\neq r,
\end{equation}
where $\Lambda_i=+1$ for senders, $\Lambda_i=0$ for transit nodes, and the receiver is treated as an unconstrained sink ($\Lambda_r=-\infty$ following the convention of Ref.~\cite{PhysRevLett.108.208701}).
As shown in Ref.~\cite{PhysRevLett.108.208701}, for an even edge cost $\phi(I)$ which increases monotonically with the flow magnitude $I$, e.g.\ $\phi(I)=|I|^\gamma$, the ground state contains no counter-flow on any edge. In this case, minimizing
\begin{equation}
H=\sum_{(ij)} \phi\!\left(\sum_{s}(\sigma^{(s\to r)}_{ij}+\sigma^{(s\to r)}_{ji})\right)
\end{equation}
is equivalent to minimizing
\begin{equation}
H'=\sum_{(ij)}\phi(I_{ij})
\end{equation}
under the flow conservation constraints $R_i=0$ for all $i\neq r$, where the sum over $(ij)$ runs over each undirected edge once. For the nonnegative edge load $x=|I|$, the cost function $\phi(x)=x^\gamma$ is concave for $0<\gamma<1$, linear at $\gamma=1$, and convex for $\gamma>1$. We therefore refer to $0<\gamma<1$ and $\gamma>1$ as the non-convex and convex regimes, respectively.
In this work, we aim to identify locally stable fixed points of the message-passing dynamics rather than finding the global optimum for every run.

\section{The cavity optimization algorithms}
\label{sec:level3}

\subsection{The original cavity algorithm without trust region}
The above model can be solved by the algorithm introduced in Ref.~\cite{PhysRevLett.108.208701}, which is based on a distributed message-passing framework derived from the cavity method in statistical physics. This algorithm computes an optimal coordinated path configuration for all senders to the receiver simultaneously, given a network topology and an arbitrary cost function. Here, we provide a summary of this message-passing algorithm:
\begin{itemize}
\item
The messages from the receiver are fixed and given by $E_{r\to j}(I_{rj})=\phi(I_{rj})$ for all $j\in \mathcal{N}_r$. 
\item
For non-receiver nodes, a message from node \(i\) to node \(l\) is defined as \(E_{i \to l}(I_{il})\), where \(I_{il}\) represents the flow from \(i\) to \(l\). Assuming that the local network structure around node $i$ is tree-like, \(E_{i \to l}(I_{il})\) denotes the minimum cost of the subtree rooted at node \(i\), given a flow \(I_{il}\) on the edge \(i \to l\). Under the flow conservation constraint, this leads to the recursive equation:
\begin{equation}
\begin{aligned}
E_{i\to l}(I_{il})
={}& \phi(I_{il}) +
\min_{\substack{
\{I_{ji}\}_{j\in\mathcal{N}_i\setminus\{l\}}\\
R_i=0}}
\sum_{j\in\mathcal{N}_i\setminus\{l\}}
E_{j\to i}(I_{ji}),
\end{aligned}
\label{eq:cavity_original}
\end{equation}
where the minimization over the set of flows \(\{I_{ji}\}\) is subject to the flow conservation constraint \(R_i = 0\) at node \(i\). 
\end{itemize}

Since the functional message $E_{i \to l}(I_{il})$ is an extensive quantity which accumulates along the tree, we define an intensive (vertex) energy as
\begin{equation}
E^V_{i\to l}(I_{il})
\equiv 
E_{i\to l}(I_{il})-E_{i\to l}(0). 
\end{equation}
The quantity $E^V_{i\to l}(I_{il})$ is then passed as the message from node $i$ to node $l$.

\subsection{Trust-region cavity algorithm}

\begin{figure}[htbp]
\centering
\includegraphics[width=\columnwidth]{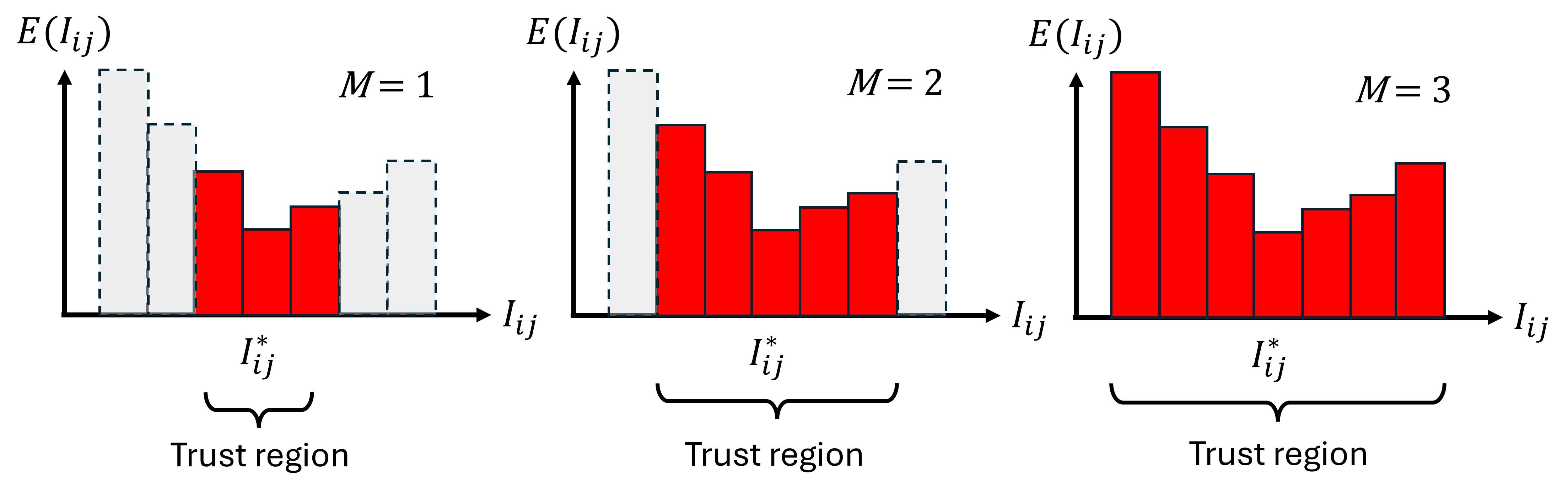}
\caption{Schematic illustration of the trust-region cavity field for different values of $M$. 
From left to right, the trust-region sizes are $M=1$, $M=2$, and $M=3$. 
The red bars represent the cavity-field values retained within the local window $m\in\{-M,\ldots,M\}$ around the current working point, while the gray dashed bars indicate values outside the trust region that are not retained.}
\label{fig:trust_region_window}
\end{figure}

With the initial goal to reduce the size of the functional-message representation, we introduce a trust-region restriction on the messages: each local update explores candidate flow changes only within a discrete window \(m\in\{-M,\ldots,M\}\) around the current working point \(I_{il}^*\), as illustrated in Fig.~\ref{fig:trust_region_window} for different values of \(M\). Unlike trust-region methods based on quadratic surrogates or Kullback--Leibler constraints, we do not introduce a second-order approximation or an additional divergence constraint. Here, the trust region is enforced directly by truncating the message domain to this local window, while the update itself is obtained by minimization restricted to this window. This truncation is facilitated by the discrete nature of the model.

The message function \(E_{i \to l}(I)\), defined over integer-valued flows \(I\), is expensive to store and transmit as a full table. Instead, we only keep its relative values within a finite window around the current working point \(I_{il}^*\) provided by the parent node \(l\). We denote the set of allowed local increments by \(\mathcal{M}\equiv\{-M,\ldots,M\}\). Hence, for each \(m\in\mathcal{M}\), we define

\begin{equation}
h^m_{i\to l}
\equiv
E^V_{i\to l}(I^*_{il}+m)
-
E^V_{i\to l}(I^*_{il}),
\label{eq:hdef}
\end{equation}

\noindent where \(h^0_{i\to l}=0\), reducing each functional message to \(2M+1\) numbers. Equivalently, within the local window, the message values satisfy

\begin{equation}
E^V_{i\to l}(I^*_{il}+m)
=
E^V_{i\to l}(I^*_{il})
+
\sum_{m'\in\mathcal{M}}
h^{m'}_{i\to l}\,\delta_{m',m}.
\label{eq:hreconstruct}
\end{equation}

Using the above recursion relations and the working-point expansion, the cavity fields are updated by

\begin{equation}
\begin{aligned}
h^m_{i\to l} \approx{}&
\phi(I^*_{il}+m)-\phi(I^*_{il}) \\
&+Q\!\left[
\{\vec h_{j\to i}\},
\{I^*_{ji}\};
\Lambda_i,I^*_{il},m
\right] \\
&-Q\!\left[
\{\vec h_{j\to i}\},
\{I^*_{ji}\};
\Lambda_i,I^*_{il},0
\right],
\end{aligned}
\label{eq:hm}
\end{equation}

\noindent where

\begin{equation}
\begin{aligned}
&Q\!\left[
\{\vec h_{j\to i}\},
\{I^*_{ji}\};
\Lambda_i,I^*_{il},m
\right] \\
&\qquad =
\min_{\substack{
\{m_j\}\\
m_j\in\mathcal{M},\;R'_i=0}}
\Bigg[
\sum_{j\in\mathcal{N}_i\setminus\{l\}}
\sum_{m'_j\in\mathcal{M}}
h^{m'_j}_{j\to i}\,
\delta_{m'_j,m_j}
\Bigg].
\end{aligned}
\label{eq:Q}
\end{equation}

The flow-conservation constraint can now be expressed as

\begin{equation}
R'_i
\equiv
\sum_{j\in\mathcal{N}_i\setminus\{l\}}
(I^*_{ji}+m_j)
+\Lambda_i-I^*_{il}-m
=0.
\label{eq:Rprime}
\end{equation}

\noindent The approximation in Eq.~\eqref{eq:hm} becomes exact as \(M\to\infty\).

In addition to updating the message \(\vec h_{i\to l}\) sent to the parent \(l\), we must update the working points of all descendants \(j\in\mathcal{N}_i\setminus\{l\}\). The minimizing increments $\{m_j^*\}$ for $j\in\mathcal{N}_i\setminus\{l\}$ are

\begin{equation}
\begin{aligned}
\{m_j^*\}
&=
\arg\min_{\substack{
\{m_j\}\\
m_j\in\mathcal{M},\;R'_i=0}}
\Bigg[
\sum_{j\in\mathcal{N}_i\setminus\{l\}}
\sum_{m'_j\in\mathcal{M}}
h^{m'_j}_{j\to i}\,
\delta_{m'_j,m_j}
\Bigg].
\end{aligned}
\label{eq:mstar}
\end{equation}

\noindent The corresponding working points are then updated according to

\begin{equation}
I^*_{ji}\leftarrow I^*_{ji}+m_j^*,
\qquad
j\in\mathcal{N}_i\setminus\{l\}.
\label{eq:working_update}
\end{equation}

\noindent For each local update at node \(i\), we enumerate all \(m_j\in\mathcal{M}\) subject to the single constraint \(R'_i=0\), evaluate \(Q\) in Eq.~\eqref{eq:Q}, update \(\vec h_{i\to l}\) using Eq.~\eqref{eq:hm}, and update the descendant working points using Eqs.~\eqref{eq:mstar} and \eqref{eq:working_update}.

In summary, by passing the messages defined in~\cite{PhysRevLett.108.208701,yeung2013efficient}, the algorithmic implementation proceeds according to the following steps:

\begin{enumerate}
\item Randomly draw a node $i$ and choose one neighbor $l\in \mathcal{N}_i$ as the parent.

\item Read the parent working point $I^*_{il}$ and gather $\{\vec h_{j\to i},I^*_{ji}\}$ for all descendants $j\in \mathcal{N}_i\setminus \{l\}$.

\item Enumerate all assignments \(\{m_j\}\) with \(m_j\in\mathcal M\), subject to $R'_i=0$ (cost $\sim(2M+1)^{|\mathcal{N}_i|-1}$) to evaluate $Q$ and obtain the minimizing increments, update $\vec h_{i\to l}$ using Eq.~\eqref{eq:hm}, and update all descendant working points $I^*_{ji}$ using Eqs.~\eqref{eq:mstar} and \eqref{eq:working_update}.

\item Repeat Steps~1--3 until convergence, i.e., no working point $I^*_{ij}$ changes in a full sweep (and equivalently the cavity fields stop changing).

\item The converged $\{I^*_{ij}\}$ define a locally optimized net-flow configuration. Each converged run yields a solution (local minimum), and the final state information (node and edge configurations) is collected for subsequent analysis.
\end{enumerate}

Each local update at node $i$ requires enumerating $(2M+1)^{k_i-1}$ assignments of $\{m_j\}$ under the single constraint $R'_i=0$, hence the per-update cost scales as $(2M+1)^{k_{\max}-1}$. Empirically, the number of updates per node before convergence grows approximately linearly with $N$, implying a total update count $\propto N^2$ and an overall complexity $O\!\left(N^2(2M+1)^{k_{\max}-1}\right)$. The approximation becomes exact as \(M \to \infty\), while small \(M\) already provides stable results in the parameter ranges examined here.

\section{Results}
\label{sec:level4}

\subsection{Identification of local minima by the trust-region cavity algorithm}

While the conventional cavity message-passing algorithms are useful for identifying solutions to optimization problems, they do not converge well when numerous local minima are present in rugged energy landscapes. Hence, variants of the message-passing algorithms, such as survey propagation~\cite{braunstein2005survey,krzakala2007gibbs}, have been developed to incorporate into the algorithm the physical insights from replica symmetry breaking (RSB)~\cite{mezard1987spin} originally derived to explain the spin-glass phase. Survey propagation was shown to perform much better than the conventional cavity approaches. Here, without applying the insights of RSB, we find that the message-passing algorithms with a trust region converge much better in non-convex optimization problems than the algorithm without a trust region.

\begin{figure}[t] 
\centering
\includegraphics[width=\columnwidth]{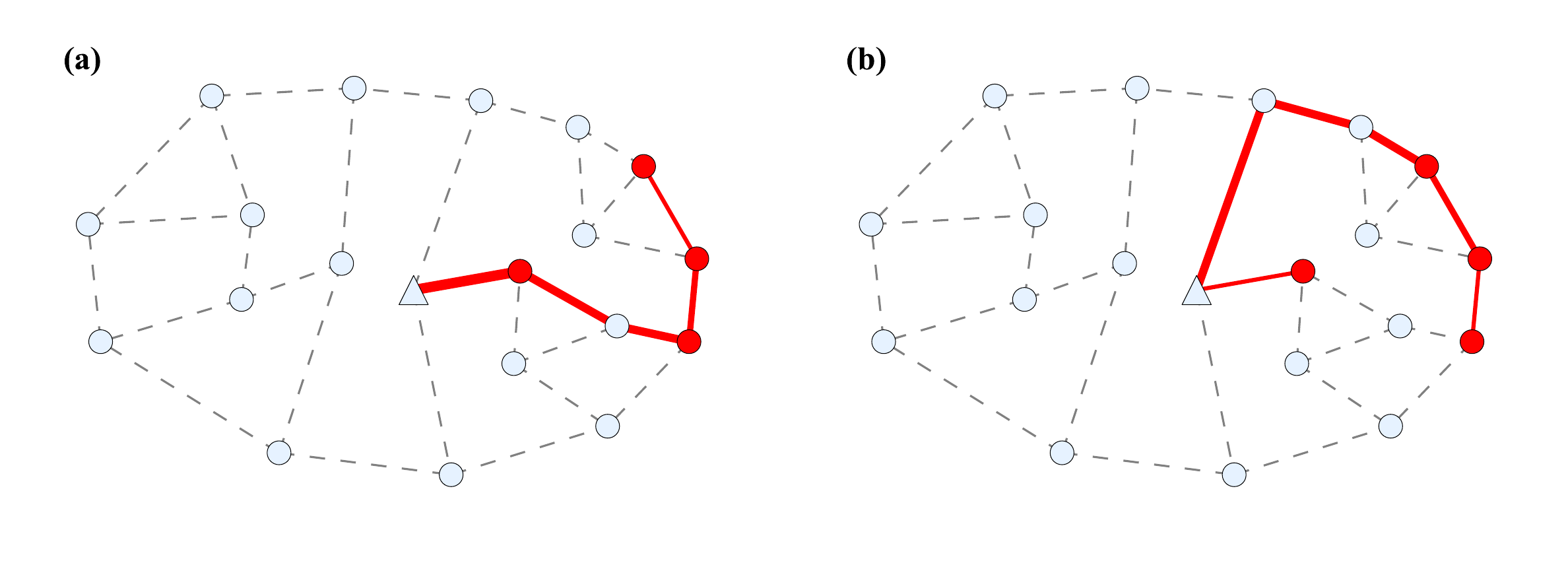} 
\caption{Two replicas illustrating different converged local minima of $H'$ with $\gamma=0.2$ on a random regular graph of $N = 20$ and $k = 3$ with a single receiver (triangle) and four senders (red circles). Red lines correspond to occupied edges, with their thickness proportional to the flow magnitude.}
\label{fig1}
\end{figure}

Specifically, we examine the regime of the model in which the energy landscape is non-convex, i.e., $\gamma < 1$. For each value of $\gamma$, multiple replicas with identical network configuration including the same network topology and the same set of senders and receiver, were simulated. Following Ref.~\cite{PhysRevLett.108.208701}, we introduce independent small random edge biases $\epsilon_{ij}$ to break degeneracy and improve convergence, i.e., the algorithm minimizes the perturbed cost $\phi(I_{ij})+|I_{ij}|\epsilon_{ij}$. While different replicas use different sets of biases to probe different local minima, the energies of the identified local minima are evaluated using the original unperturbed cost $\phi(I)$ without the biases for comparison across replicas. As an example, Fig.~\ref{fig1} illustrates the different converged states identified by the trust-region cavity algorithm, which correspond to physically distinct path configurations.

\subsubsection*{(i) Verification of the identified solutions as local minima}

While we expect that the solutions identified by the trust-region cavity algorithm correspond to global or local minima, there is a possibility that some of them are only suboptimal path configurations produced by the finite information retained within the trust region. Before performing explicit neighborhood-expansion tests, we first examine whether the identified states behave consistently with the expected change from the non-convex to the convex regime. 

Figure~\ref{fig:landscape_diagnostics} combines three complementary diagnostics for the representative sender fraction $f_p=0.5$: (a) the number of distinct converged energy values, (b) the convergence rate, and (c) the unique-energy ratio among converged replicas. In the convex regime $\gamma>1$, Fig.~\ref{fig:landscape_diagnostics}(a) shows that all tested trust-region sizes $M$ identify only a single energy value, while Fig.~\ref{fig:landscape_diagnostics}(b) shows convergence approaching unity and Fig.~\ref{fig:landscape_diagnostics}(c) shows the corresponding unique-energy ratio becoming very small. This provides a baseline consistency check that the trust-region cavity algorithm repeatedly reaches the same energy level when the landscape is expected to be dominated by a unique minimum, even for a small trust region such as $M=1$.

By contrast, for $\gamma<1$, Fig.~\ref{fig:landscape_diagnostics}(a) shows multiple distinct converged energy values, consistent with the emergence of a non-convex landscape containing multiple minima. To verify that the identified solutions in this regime are locally stable rather than artifacts of an insufficient trust region, we examine their response to an increase in $M$ after the messages have converged. As a first verification, after all messages in the network converge, we randomly selected one edge and increase $M$ for the messages associated with that edge. In other words, we do not directly re-route the selected edge, but expand its local trust region and allow the algorithm to warm-start from the originally identified solution and re-converge.

\begin{figure}[t]
\centering
\includegraphics[width=\columnwidth]{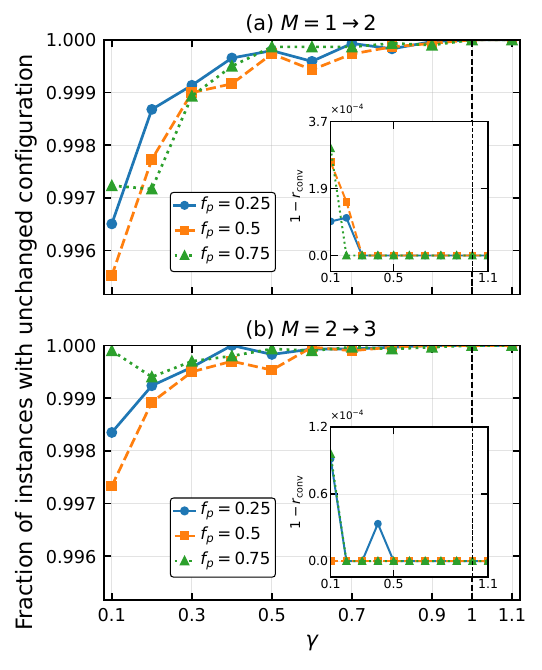}
\caption{Chained single-edge neighborhood-expansion stability test.
After convergence of the message-passing algorithm, one edge is randomly selected and its local window is enlarged successively from $M=1$ to $M=2$ and, after re-convergence, from $M=2$ to $M=3$. The two main panels show the fraction of unchanged outcomes, defined by $|\Delta E|\leq\epsilon$ with $\Delta E=E_{\mathrm{new}}-E_{\mathrm{old}}$, for the two stages of the chained protocol. The upper panel corresponds to the $M=1\to2$ expansion, and the lower panel corresponds to the subsequent $M=2\to3$ expansion applied to the same selected edge. Different curves represent different sender fractions $f_p$. The consistently high unchanged fraction indicates that the converged fixed points are stable under successive single-edge neighborhood enlargement. Insets show the corresponding post-expansion non-convergence fraction, $1-r_{\mathrm{conv}}$, where $r_{\mathrm{conv}}$ is the post-expansion convergence rate. The small values of $1-r_{\mathrm{conv}}$ confirm that failed post-expansion convergence is rare and that the observed stability is not dominated by non-converged runs.}
\label{fig:chain_edge_outcomes}
\end{figure}

We considered three protocols: (1) the trust region of the messages on a selected edge was increased from $M=1$ to $M=2$; (2) the trust region of the messages on a selected edge was increased from $M=2$ to $M=3$; (3) a chained test in which the trust region of the same selected edge was first expanded from $M=1$ to $M=2$, and if the messages of the whole network re-converged, the trust region was further expanded from $M=2$ to $M=3$. The chained protocol provides a stronger probe because it tests whether the identified solutions remain stable under successive increases in the trust region. We denote the unperturbed objective value $H'$ of a converged state by $E$. The energy change after the neighborhood expansion is then $\Delta E = E_{\rm new} - E_{\rm old}$. $\Delta E$ can be zero, negative, or positive, corresponding to a new solution with equal, lower, or higher energy, respectively. As shown in Fig.~\ref{fig:chain_edge_outcomes}, the unchanged outcome dominates in the chained protocol, suggesting that the identified solutions are locally stable. The corresponding independent single-step tests are reported in Appendix~\ref{app:edge_expansion_robustness}.

\subsubsection*{(ii) Local stability under single-edge neighborhood expansion}
Figure~\ref{fig:chain_edge_outcomes} summarizes the unchanged fractions for the chained test. The upper panel corresponds to the first expansion, $M=1\to2$, and the lower panel corresponds to the second expansion, $M=2\to3$, applied to the same selected edge. Across both stages, the unchanged outcome dominates for every tested value of $\gamma$ and all sender fractions $f_p$. In contrast, energy-changing outcomes remain very rare. Thus, a converged fixed point is rarely destabilized by enlarging the local neighborhood of a randomly selected edge, even when the local neighborhood of that same edge is enlarged successively.

To verify that the dominance of the unchanged category is not caused by failed post-expansion convergence, we also measured the post-expansion convergence rate. The insets of Fig.~\ref{fig:chain_edge_outcomes} show the corresponding post-expansion non-convergence fraction, which remains small over the tested parameter range in both stages of the chained protocol. Consequently, the small fraction of energy-changing events is not an artifact of failed re-convergence after the local expansion, but instead reflects the stability of the original fixed points under the tested neighborhood enlargement.

Across the chained protocol and the two independent single-step protocols reported in Appendix~\ref{app:edge_expansion_robustness}, the overwhelming majority of converged fixed points remain unchanged. Both energy-decreasing and energy-increasing events are rare, while the post-expansion convergence rate remains high. The results indicate that the fixed points obtained by the truncated message-passing algorithm are not merely artifacts of an artificially narrow local window; within the tested range of neighborhood enlargements, they behave as locally stable low-energy states.

This procedure is a numerical stability test rather than a formal proof of local optimality. Because the neighborhood of only one randomly selected edge is expanded at a time, it does not exclude every possible coordinated multi-edge rearrangement. Nevertheless, the consistent behavior across the $M=1\to2$, $M=2\to3$, and chained $M=1\to2\to3$ protocols provides strong evidence that the converged message-passing states are robust local low-energy configurations.

\subsection{The characteristics of local minima}
\begin{figure}[t]
\centering
\includegraphics[width=\columnwidth]{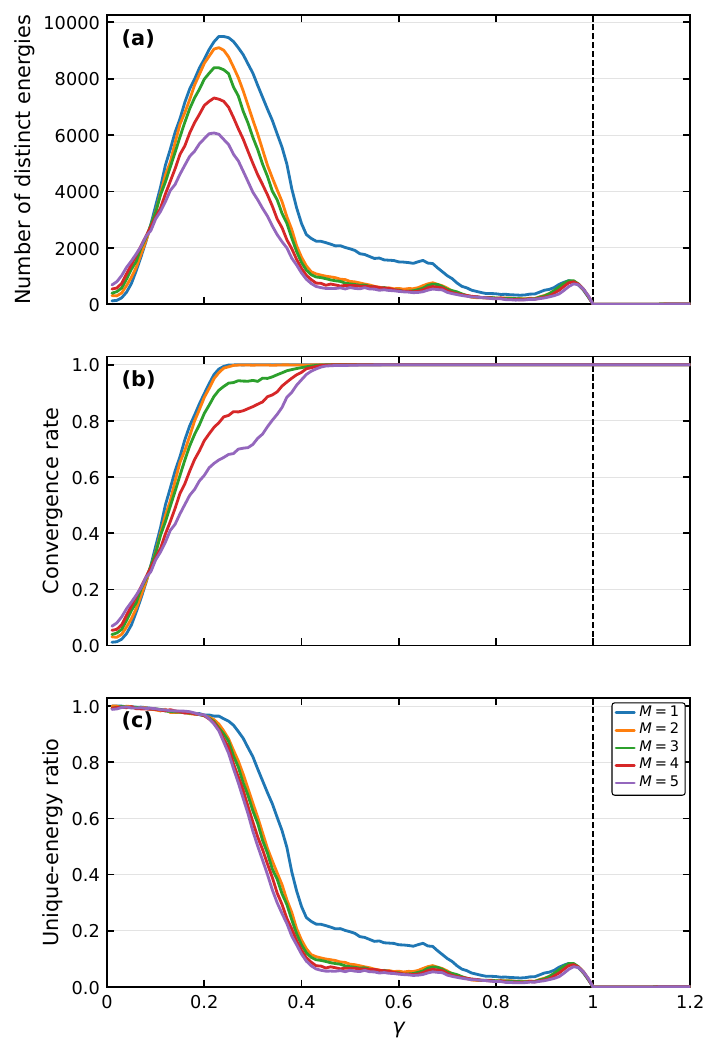}
\caption{Basic landscape diagnostics of the identified local minima for the representative sender fraction $f_p=0.5$.
(a) Number of distinct converged energy values, (b) convergence rate, and (c) unique-energy ratio among converged replicas as functions of $\gamma$ for $M=1,2,3,4,5$. The dashed vertical line marks $\gamma=1$, separating the non-convex and convex regimes considered in the text.}
\label{fig:landscape_diagnostics}
\end{figure}

Given that the solutions identified by the trust-region cavity algorithm are likely to be minima, we further examine their basic landscape characteristics using the three diagnostics collected in Fig.~\ref{fig:landscape_diagnostics}. Since the three tested sender fractions exhibit qualitatively similar trends in these diagnostics, we present $f_p=0.5$ as a representative case for clarity. Figure~\ref{fig:landscape_diagnostics}(a) shows the number of distinct converged energy values as a function of $\gamma$ for $M=1,\ldots,5$. In the non-convex regime $\gamma<1$, this number exhibits a pronounced maximum near $\gamma\approx0.2$ across different $M$ values, indicating the largest diversity of identified low-energy states around this parameter range. We remark that the number of distinct flow configurations would more directly represent the number of local minima; here, however, we use the number of distinct converged energy values as a proxy because the difference between these two quantities is small in our parameter range.

\begin{figure}[htbp]
\centering
\includegraphics[width=\columnwidth]{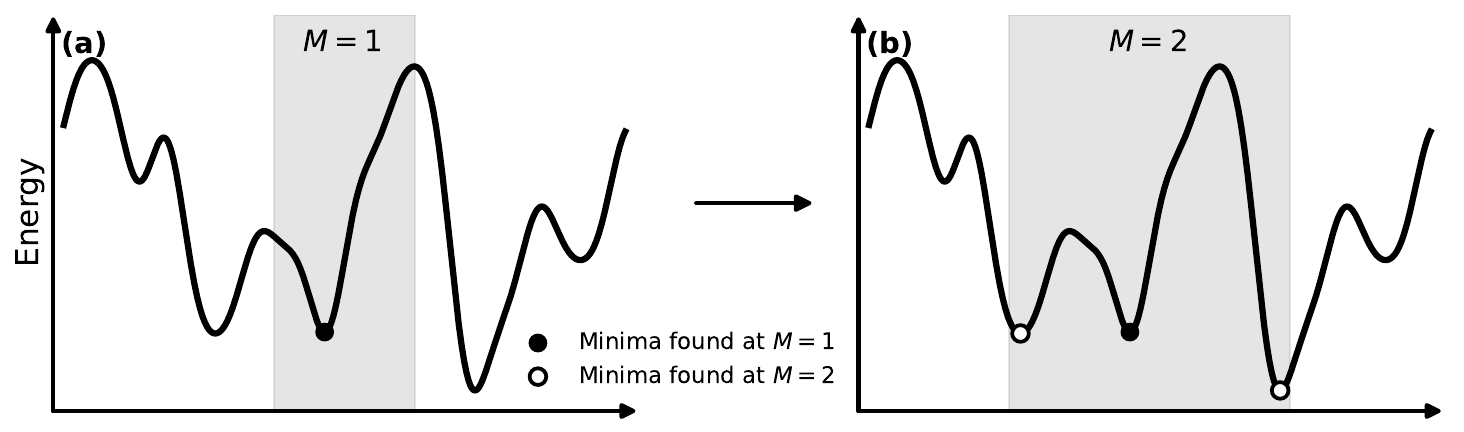}
\caption{Schematic illustration of the effect of enlarging the trust region.
The two panels show the same schematic one-dimensional non-convex energy landscape. 
For $M=1$, the narrower trust region restricts the accessible local search range and the algorithm identifies a local minimum within the corresponding neighborhood. 
When the trust region is enlarged to $M=2$, a broader region of the landscape becomes accessible, allowing the algorithm to identify other local minima outside the narrower neighborhood. 
The filled and open circles denote the minima identified at $M=1$ and $M=2$, respectively.}
\label{fig:trust_region_schematic}
\end{figure}

Figure~\ref{fig:landscape_diagnostics}(b) shows that the algorithm's convergence rate depends strongly on both $\gamma$ and $M$. For the representative case $f_p=0.5$, the convergence rate is generally highest when $M=1$. This is intriguing, as a smaller trust region, i.e., a smaller $M$, means less local information is provided to the algorithm, and hence it explores a smaller region of the complex energy landscape in the non-convex regime. As schematically illustrated in Fig.~\ref{fig:trust_region_schematic}, for small $M$, the algorithm can only explore the local region of the valley of the energy landscape toward which it is currently converging, whereas increasing $M$ enlarges the accessible region and may allow the algorithm to identify another local minimum. Hence, the restricted exploration at small $M$ may reduce the tendency of the algorithm to move toward another local minimum, thereby improving its convergence. This is a counterintuitive but important implication of applying a trust region in the message-passing algorithm, which facilitates both convergence and the identification of local minima. 

Although we suggest that the higher convergence rate at small $M$ results from the smaller region accessible to the algorithm in rugged energy landscapes, one may wonder whether the observed non-convergence is instead related to the loopy network topology, given that cavity message-passing algorithms are not expected to converge well on loopy graphs. To show that loopy graphs are not the cause of the non-convergence, we note from Fig.~\ref{fig:landscape_diagnostics}(b) that convergence is perfect on the same loopy networks when $\gamma>1$, regardless of $M$, where the energy landscape is convex and characterized by a single global minimum. Non-convergence occurs only when $\gamma<1$, i.e., when the energy landscapes are non-convex. This observation rules out the loopy network topology as the origin of the non-convergence observed for $\gamma<1$ and suggests that the improved convergence at small $M$ may instead be related to the interplay between $M$ and the ruggedness of the energy landscapes.

Because the absolute count in Fig.~\ref{fig:landscape_diagnostics}(a) depends on how many replicas actually converge, Fig.~\ref{fig:landscape_diagnostics}(c) provides a complementary normalized diagnostic through the unique-energy ratio among converged replicas. For $f_p=0.5$, this ratio decreases overall as $\gamma$ increases. In particular, the marked decrease beginning near $\gamma\approx0.2$ is consistent with the diversity peak in Fig.~\ref{fig:landscape_diagnostics}(a), while the ratio becomes very small as $\gamma$ approaches and enters the convex regime. A minor secondary feature is also visible near $\gamma\approx1$. Taken together, the three panels in Fig.~\ref{fig:landscape_diagnostics} show that increasing $\gamma$ is accompanied by fewer distinct identified energies, higher convergence, and a progressively smaller fraction of unique energies among the converged replicas.

\begin{figure*}[!t]
\centering
\includegraphics[width=\linewidth]{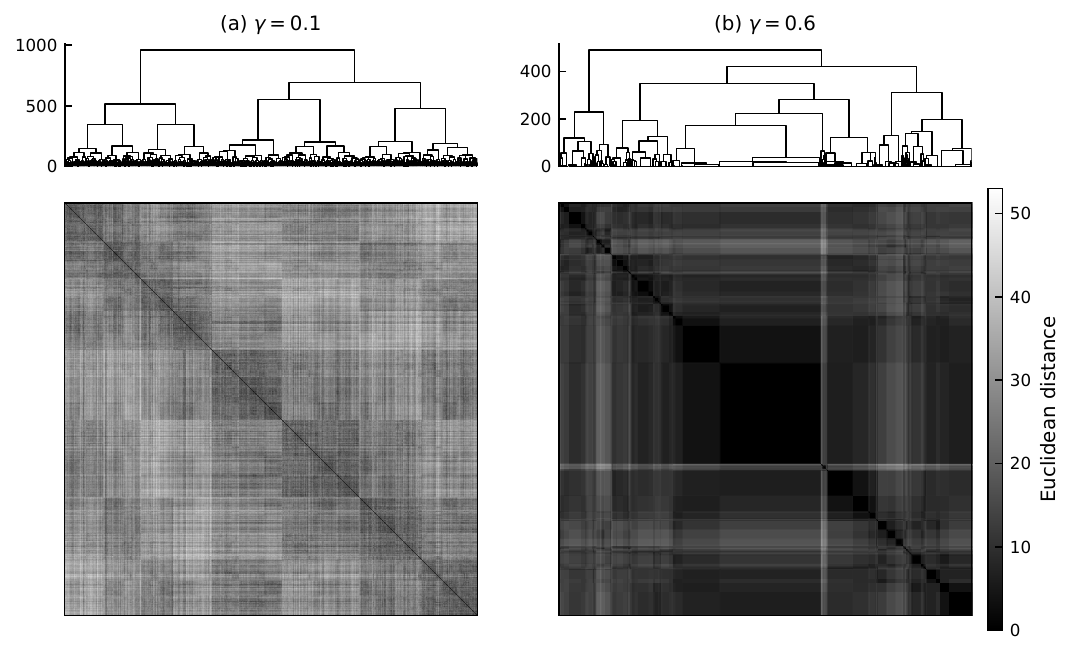}
\caption{Hierarchical clustering dendrograms and reordered distance-matrix heatmaps for $N=100$, $k=3$, and $f_p=0.25$, with (a) $\gamma=0.1$ and (b) $\gamma=0.6$.}
\label{fig5}
\end{figure*}

Since the trust-region cavity algorithm can identify a set of local minima, we can analyze their hierarchical structure using methods similar to those applied to ground states in spin glasses~\cite{hed2001spin}. For the non-convex regime with $\gamma<1$, we collect the edge-flow configurations from 10,000 converged solutions obtained on the same network instance with $f_p=0.25$ and compute the Euclidean distance between two solutions $a$ and $b$,

\begin{equation}
d_{ab}
=
\left[
\sum_{\ell=1}^{N_e}
\left(I_\ell^{(a)}-I_\ell^{(b)}\right)^2
\right]^{1/2},
\end{equation}

\noindent where $I_\ell^{(a)}$ and $I_\ell^{(b)}$ denote the flows on edge $\ell$ in solutions $a$ and $b$, respectively, and $N_e$ is the number of edges in the network.
We then apply Ward's hierarchical clustering~\cite{ward1963hierarchical}, resulting in the dendrograms and reordered distance matrices shown in Fig.~\ref{fig5}, where darker shades indicate smaller distances between solutions.

Figure~\ref{fig5}(a), for $\gamma=0.1$, reveals a coarse-grained structure in which the solution space partitions into several clusters. Although these clusters are well-separated in the dendrogram, the relatively light color of the heatmap indicates considerable distances between configurations, and the cluster structure is not as clear as that in the dendrogram. The recurring gray blocks within clusters reveal several subclusters, signifying a hierarchical organization resembling that observed in spin glasses~\cite{PhysRevLett.86.3148}. The dendrogram and heatmap in Fig.~\ref{fig5}(b) for $\gamma=0.6$ show qualitative difference from those for $\gamma=0.1$. The emergence of numerous dark blocks within clusters indicates that configurations form tightly packed groups, revealing a hierarchical structure.

Taken together, these landscape diagnostics suggest three qualitatively different regimes as $\gamma$ increases. Around $\gamma\approx0.2$, the diversity of identified low-energy states reaches a maximum, corresponding to the most rugged region of the landscape. For $0.2\lesssim\gamma<1$, the diversity of identified minima decreases, while the low-energy solution space retains a hierarchical multi-cluster organization. For $\gamma>1$, the identified energies collapse toward a single value and the landscape becomes dominated by a unique or nearly unique minimum. These regimes should be understood as qualitative crossovers in the observed landscape structure rather than sharply defined thermodynamic phases.

\subsection{Probing the energy landscape by increasing the size of the trust region}
As illustrated in Fig.~\ref{fig:trust_region_schematic}, we can obtain more information about the energy landscape by progressively increasing the size of the trust region. Hence, we start from local minima identified with $M=1$ and progressively expand the trust region to $M=5$ to probe whether additional local minima become accessible. For every consecutive expansion for which the algorithm successfully re-converges, we record the change in energy as $\Delta E_{M\to M+1}$ and pool the outcomes of all consecutive trust-region expansions for each $(\gamma,f_p)$ into three categories: \texttt{Decrease} ($\Delta E<0$), \texttt{Unchanged} ($\Delta E=0$), and \texttt{Increase} ($\Delta E>0$) (Fig.~\ref{fig_incdec}). We also report (i) the decrease-to-increase ratio among nontrivial changes, $f_{\mathrm{DI}}=N_{\mathrm{dec}}/N_{\mathrm{inc}}$, where $N_{\mathrm{dec}}$ and $N_{\mathrm{inc}}$ are the numbers of energy-decreasing and energy-increasing outcomes, respectively (Fig.~\ref{fig7}(a)), and (ii) the re-convergence fraction, $r_{\mathrm{conv}}=N_{\mathrm{conv}}/40000$, where $40000=1000\times10\times4$ corresponds to 1000 samples, 10 replicas per sample, and four adjacent $M$-expansion steps (Fig.~\ref{fig7}(b)).

A first observation is that for $\gamma\ge 1$, enlarging $M$ produces no change, consistent with the fact that the landscape is dominated by a unique minimum in this convex regime, and the algorithm does not identify a different-energy minimum within the tested range of trust-region sizes. On the other hand, for $\gamma<1$, the stability of the $M=1$ solution depends on $f_p$. When $f_p=0.25$, the \texttt{Unchanged} fraction increases markedly with $\gamma$, and the total probability mass of \texttt{Decrease} and \texttt{Increase} shrinks accordingly. In contrast, for $f_p=0.5$ and $0.75$, the \texttt{Unchanged} fraction also increases with $\gamma$ but with a noticeably weaker dependence, and nonzero fractions of \texttt{Decrease} and \texttt{Increase} persist over a broader range of $\gamma<1$.

\begin{figure}[t]
\centering
\includegraphics[width=\columnwidth]{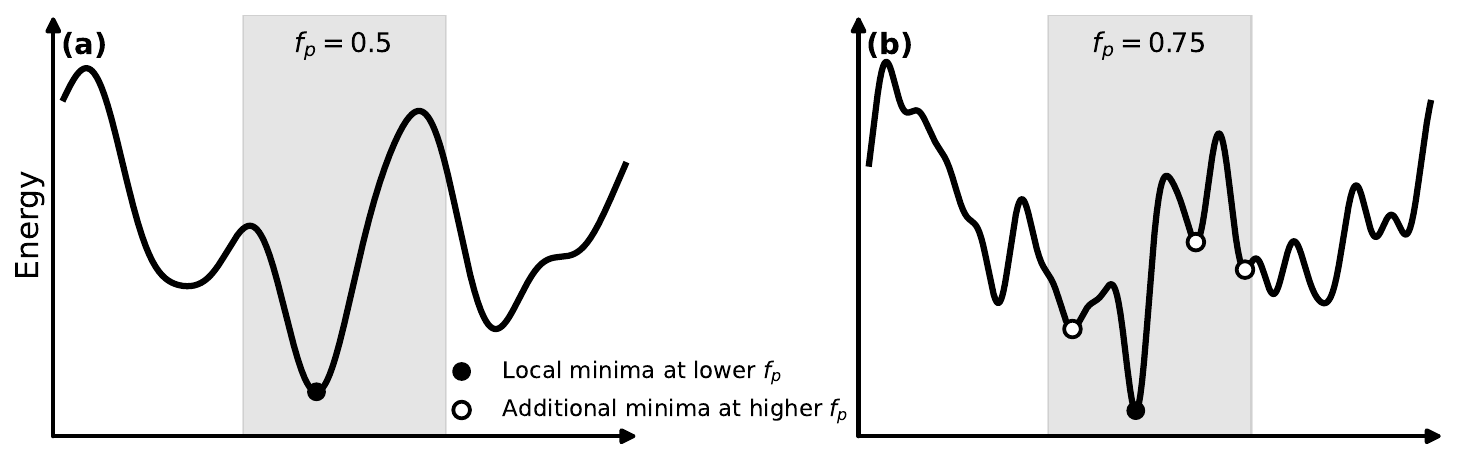}
\caption{Schematic illustration of how the response to trust-region expansion reflects the local ruggedness of the energy landscape. (a) A less rugged local landscape, representative of the case with $f_p=0.5$, in which the fixed-width trust region (gray region) covers essentially a single broad valley, so enlarging the trust region is less likely to identify a different local minimum. (b) A more rugged local landscape, representative of the case with $f_p=0.75$, in which the same trust-region width covers multiple narrower nearby valleys, so enlarging the trust region can more easily identify additional local minima. The filled circle denotes the current local minimum, and the open circles denote other local minima accessible within the same trust region.}
\label{fig:ruggedness_schematic}
\end{figure}

\begin{figure}[t] 
\centering
\includegraphics[width=\columnwidth]{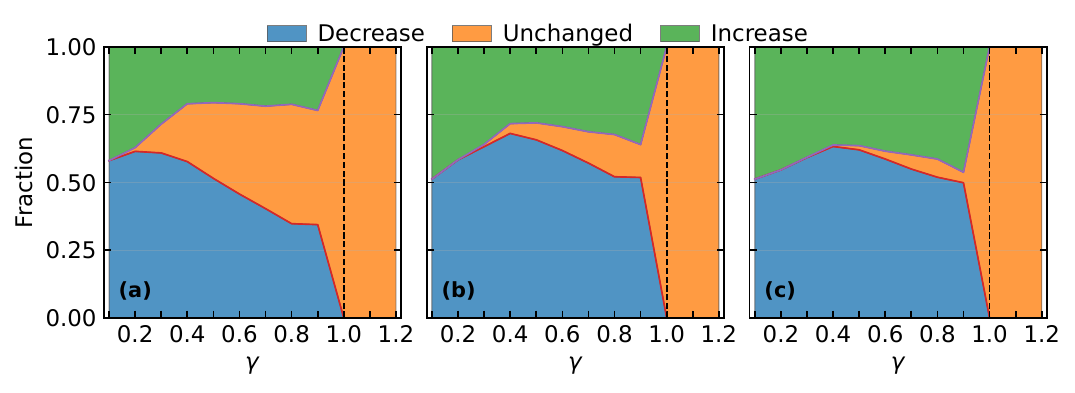} 
\caption{Stacked proportions of energy-change outcomes induced by enlarging the trust-region width from $M=1$ to $M=5$. For each $(\gamma, f_p)$, all successful adjacent-stage comparisons ($M=1\!\to\!2$, $2\!\to\!3$, $3\!\to\!4$, $4\!\to\!5$) are pooled and categorized by the sign of the unperturbed-energy change $\Delta E_{M\to M+1}$: \texttt{Increase} ($\Delta E>0$), \texttt{Decrease} ($\Delta E<0$), and \texttt{Unchanged} ($\Delta E=0$). Panels (a), (b), (c) correspond to $f_p=0.25$, $0.5$, $0.75$, respectively; the dashed vertical line marks $\gamma=1$, at and above which the outcome becomes entirely \texttt{Unchanged} (including $\gamma=2$).}
\label{fig_incdec}
\end{figure}

\begin{figure}[t]
\centering
\includegraphics[width=\columnwidth]{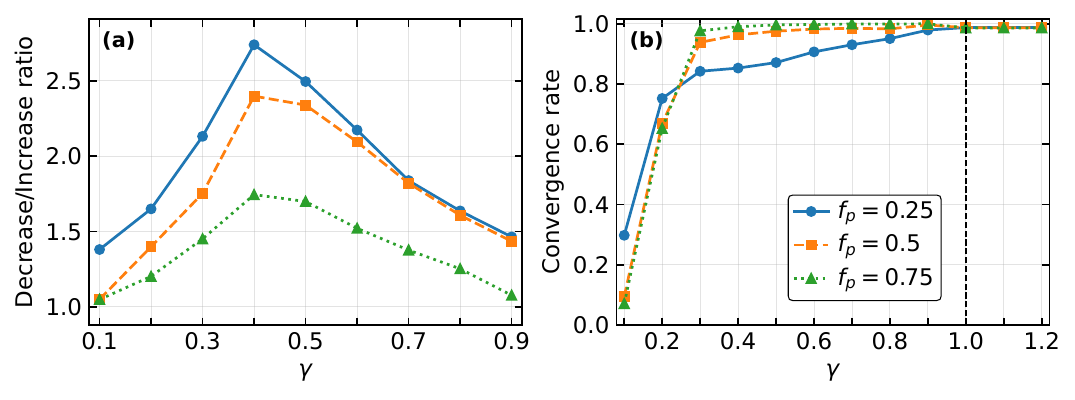}
\caption{Decrease-to-increase ratio and convergence for different $(\gamma, f_p)$.
(a) Decrease-to-increase ratio of unperturbed-energy changes during $M$ expansion.
(b) Re-convergence of successful trust-region expansions under progressive trust-region expansion.}
\label{fig7}
\end{figure}

These trends also provide qualitative information about the local structure of the energy landscape. A high \texttt{Unchanged} fraction means that enlarging the trust region is unlikely to move the system away from the originally identified minimum, suggesting that the surrounding basin is locally robust under the tested neighborhood expansion. By contrast, a lower \texttt{Unchanged} fraction indicates that the same expansion more readily reaches a different low-energy state, which is consistent with a more rugged or more finely structured landscape containing more nearby minima. In particular, compared with the case $f_p=0.5$, the case $f_p=0.75$ exhibits a smaller \texttt{Unchanged} fraction over a broader range of $\gamma<1$, suggesting that its low-energy landscape is locally more rugged under the same trust-region expansion. This interpretation is illustrated schematically in Fig.~\ref{fig:ruggedness_schematic}, where the same trust-region width covers more nearby valleys in a more rugged landscape.

Conditioned on adjacent trust-region expansions that produce a nonzero energy change ($\Delta E\neq 0$), the ratio of energy-decreasing to energy-increasing outcomes is non-monotonic in $\gamma$. For $f_p=0.25$, $f_{\mathrm{DI}}$ first rises and then falls as $\gamma$ increases toward 1 (Fig.~\ref{fig7}(a)); a qualitatively similar ``increase-then-decrease'' trend is also visible for $f_p=0.5$ and $0.75$. Thus, as $\gamma$ grows (while remaining below 1), the system initially exhibits a increasing predominance of energy-decreasing moves among the changes that occur, but this predominance weakens again as the \texttt{Unchanged} outcome becomes increasingly dominant as $\gamma$ approaches $1$. The re-convergence fraction after consecutive trust-region expansions also increases with $\gamma$ across all three $f_p$ values.

\subsection{Non-uniform trust-region sizes}

In all previous analyses, the same trust-region size is applied network-wide; here, we study a heterogeneous setting with non-uniform trust-region sizes across the network. If we consider the size of the trust region as representing the amount of information used in the optimization, and hence as a kind of resource, we can compare heterogeneous and homogeneous trust-region assignments under the condition that the total resource budget, i.e., the sum of trust-region sizes over all nodes, is the same.

\begin{figure}[t]
\centering
\includegraphics[width=\columnwidth]{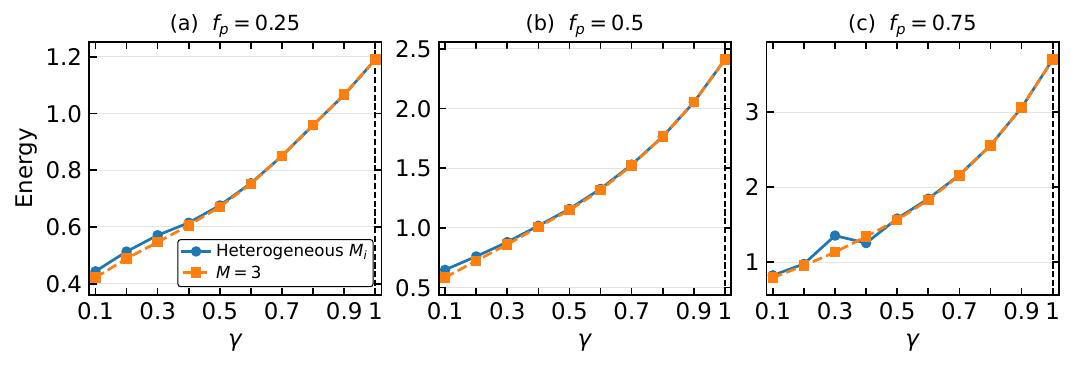}
\caption{Robustness to spatially heterogeneous trust-region size. Final energy $E$ versus $\gamma$ for three sender fractions: (a) $f_p=0.25$, (b) $f_p=0.5$, and (c) $f_p=0.75$. The heterogeneous setting assigns $M_i\in\{1,2,3,4,5\}$ across the $N=100$ nodes with 20 nodes per value, while the homogeneous baseline uses $M=3$ for all nodes. The nearly overlapping energy curves indicate that the observed landscape phenomenology is not sensitive to moderate spatial heterogeneity in the trust-region size.}
\label{fig:heteroM_vs_M3}
\end{figure}

Specifically, we randomly divide all nodes into five equally sized groups and assign trust-region sizes $M=1,\dots,5$ to the five groups, respectively, and compare this case with the homogeneous case in which all nodes use $M=3$. As shown in Fig.~\ref{fig:heteroM_vs_M3}, the dependence of the energy on $\gamma$ is essentially unchanged under heterogeneous $M_i$, and the identified energies remain close to those of the homogeneous $M=3$ case across the tested $\gamma$ range. These results suggest that spatially redistributing the trust-region sizes while keeping the same mean does not unlock lower-energy solutions.

\section{Summary}
\label{sec:level5}

This study incorporates the concept of a trust region into the cavity approach and shows that the trust-region cavity algorithm achieves higher convergence rates than cavity algorithms using full functional messages when identifying local minima in non-convex energy landscapes. By studying a route coordination problem with a single parameter controlling the convexity of the cost function, we apply the trust-region cavity algorithm to identify local minima and probe the low-energy solution space in the non-convex regime. 

The identified fixed points are further examined using progressive-$M$, heterogeneous-$M_i$, and single-edge neighborhood-expansion tests. The progressive-$M$ test reveals that, particularly in the non-convex regime, enlarging the trust region can make additional nearby minima accessible, indicating a nontrivial local landscape structure. By contrast, the heterogeneous-$M_i$ test and the single-edge neighborhood-expansion tests show that the converged energies and fixed points are generally robust to the tested spatial redistribution or localized enlargement of the trust region. Moreover, the observed changes do not lead to systematic energy reduction. Taken together, these results support the interpretation that the identified states represent local low-energy configurations rather than artifacts of an artificially restricted trust region.

Our results reveal three qualitatively different regimes of the energy landscape as $\gamma$ increases. Around $\gamma\approx0.2$, the diversity of identified local minima reaches a maximum, corresponding to a highly rugged landscape. For $0.2\lesssim\gamma<1$, the diversity of identified minima decreases and the low-energy solutions exhibit a hierarchical multi-cluster organization. For $\gamma>1$, the landscape becomes smooth and is dominated by a unique or nearly unique minimum. Overall, these observations suggest a crossover from a densely populated rugged landscape, through an intermediate hierarchical regime, to a smooth landscape dominated by a single solution.

\begin{acknowledgments}
The work is supported by the Research Grants Council of the Hong Kong Special Administrative Region, China (Projects No. GRF 18300623), and the Dean’s Research Fund of the Faculty of Liberal Arts and Social Sciences, The Education University of Hong Kong, Hong Kong Special Administrative Region, China (Projects No. FLASS/DRF 04847, DRF 0406G).
We wish to acknowledge the support of The Education University of Hong Kong for funding and computational resources. 
\end{acknowledgments}

% =========================
% Appendices
% =========================

\appendix

% -------------------------------------------------

\section{Baseline sensitivity under uniform $M$ (Mono-$M$)}
\label{app:A1_monoM}

This appendix provides supplementary diagnostics for the baseline setting in which all nodes share the same trust-region window size $M$.

The main text adopts a fixed $M$ (e.g., $M=3$) as a representative operating point; here, we show the full sweep over $M=1,2,3,4,5$ to (i) demonstrate that the reported energy-landscape phenomenology is not an artifact of one particular $M$ and (ii) justify the use of a moderate $M$ as a practical choice.

Across all tested sender fractions $f_p\in\{0.25,0.5,0.75\}$ and the $\gamma$ values reported in the main text, the sample-averaged energies vary only weakly with $M$. This indicates that, under the present algorithm, moderate values of $M$ already lead to converged states with energies close to those obtained at larger $M$, so further enlarging the local search window produces only small changes in the final energy.

\begin{figure}[b]
\centering
\begin{subfigure}{0.99\linewidth}
\centering
\includegraphics[width=\linewidth]{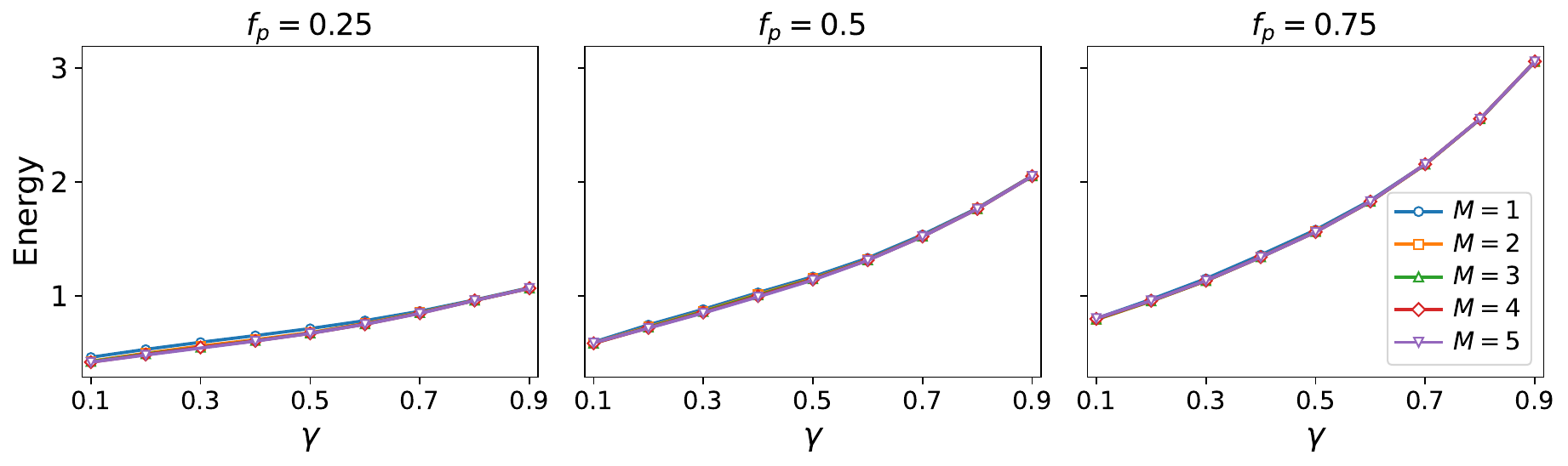}
\caption{Energy. Sample-averaged energy versus $\gamma$ for uniform
$M\in\{1,2,3,4,5\}$ at $f_p\in\{0.25,0.5,0.75\}$.}
\label{fig:app_monoM_energy}
\end{subfigure}
\hfill
\begin{subfigure}{0.99\linewidth}
\centering
\includegraphics[width=\linewidth]{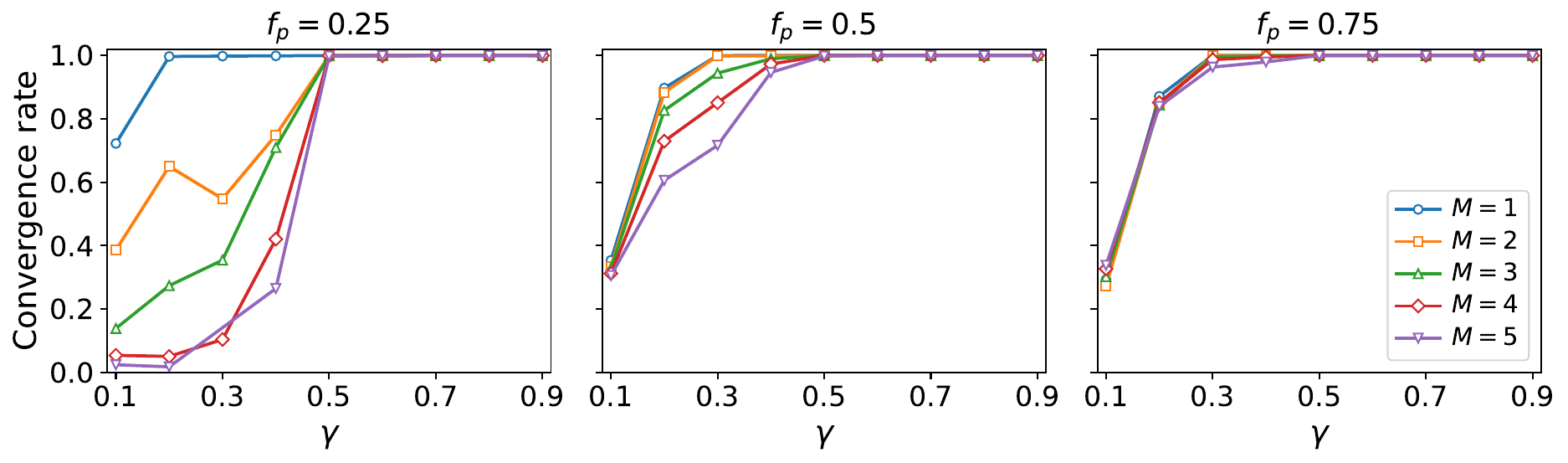}
\caption{Convergence rate. Fraction of converged replicas versus $\gamma$
for the same runs.}
\label{fig:app_monoM_conv}
\end{subfigure}
\caption{Supplementary baseline plots for uniform $M$ (Mono-$M$).
All other settings are the same as in the main text. The results show that the
converged energies are only weakly sensitive to $M$, with moderate values of
$M$ producing energies close to those obtained at larger trust-region sizes.}
\label{fig:app_monoM}
\end{figure}

% -------------------------------------------------

\section{Robustness checks for the single-edge neighborhood-expansion test}
\label{app:edge_expansion_robustness}

In the main text, we report the chained single-edge expansion test, in which the trust region associated with the same randomly selected edge is successively enlarged from $M=1$ to $M=2$ and then from $M=2$ to $M=3$. To confirm that the observed stability is not specific to the chained protocol, we also consider two independent single-step tests. In the first, the system is initialized from a fixed point obtained at $M=1$, and the trust region associated with one randomly selected edge is enlarged to $M=2$. In the second, the system is initialized from a fixed point obtained at $M=2$, and the trust region associated with one randomly selected edge is enlarged to $M=3$.

\begin{figure}[t]
\centering
\includegraphics[width=\linewidth]
{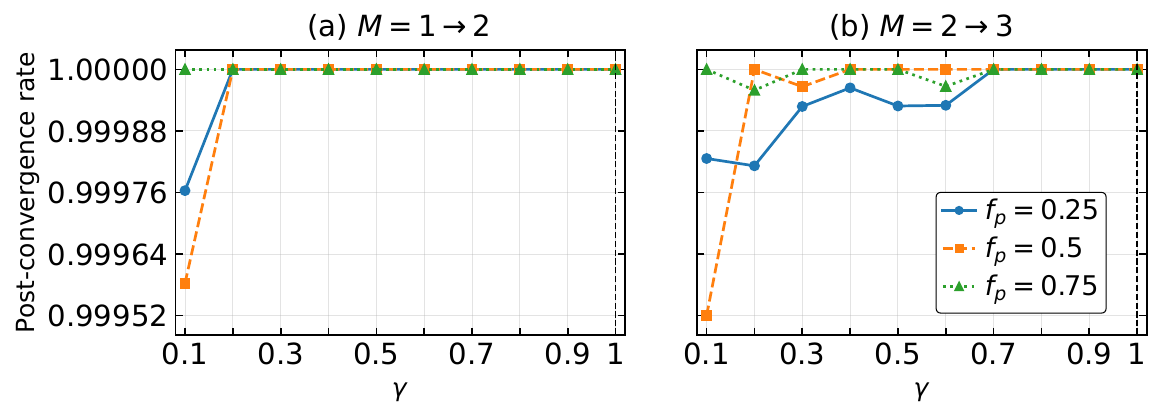}
\caption{Post-expansion re-convergence rates for the independent single-step neighborhood-expansion tests. (a) After convergence at $M=1$, the trust region associated with one randomly selected edge is enlarged to $M=2$. (b) After convergence at $M=2$, the trust region associated with one randomly selected edge is enlarged to $M=3$. In both cases, the message-passing algorithm is warm-started from the original fixed point, and the post-expansion re-convergence rate remains high across the tested values of $\gamma$ and $f_p$.}
\label{fig:app_edge_post_conv}
\end{figure}

\begin{figure*}[!t]
\centering
\includegraphics[width=0.9\textwidth]{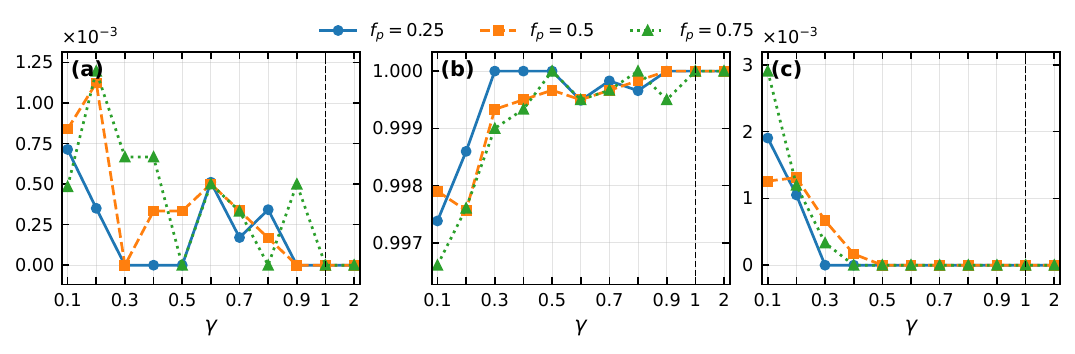}
\vspace{4mm}
\includegraphics[width=0.9\textwidth]{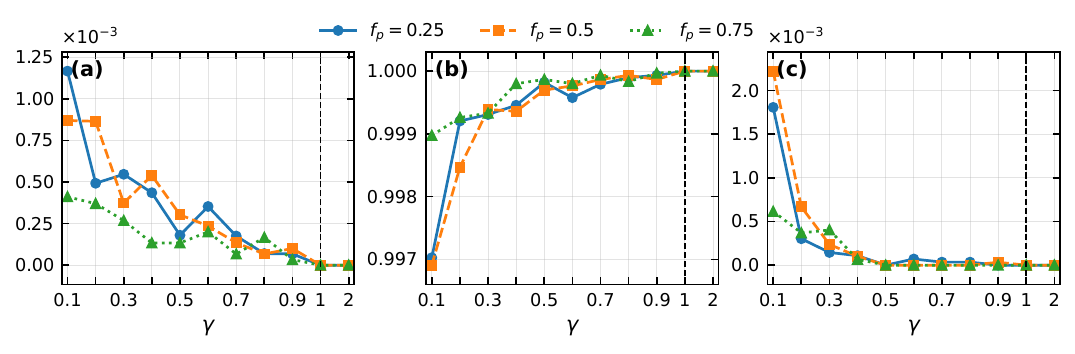}
\caption{Independent single-step neighborhood-expansion tests. The upper panels correspond to the expansion from $M=1$ to $M=2$, and the lower panels correspond to the expansion from $M=2$ to $M=3$. Within each row, panels (a), (b), and (c) show the fractions of energy-decreasing, unchanged, and energy-increasing outcomes, respectively, classified by $\Delta E=E_{\mathrm{new}}-E_{\mathrm{old}}$. For both expansions, the unchanged outcome dominates across the tested values of $\gamma$ and $f_p$, consistent with the chained test reported in the main text.}
\label{fig:app_edge_single_step}
\end{figure*}

As shown in Fig.~\ref{fig:app_edge_single_step}, the independent tests are consistent with the chained result. For both $M=1\to2$ and $M=2\to3$, the unchanged outcome dominates over the energy-decreasing and energy-increasing outcomes. The corresponding post-expansion re-convergence rates are shown in Fig.~\ref{fig:app_edge_post_conv} and remain high across the tested parameter range. These checks support the conclusion that the message-passing fixed points are stable under controlled single-edge neighborhood expansion and that the main-text observation is not peculiar to the chained protocol.

% \nocite{*}

\clearpage
\bibliography{apssamp}% Produces the bibliography via BibTeX.

\end{document}